\documentclass[prd,twocolumn, showpacs,nofootinbib,preprintnumbers]{revtex4}

\usepackage{amsmath}
\usepackage{amsfonts}
\usepackage{graphicx}
\usepackage{dcolumn}
\usepackage{hyperref}

\newcommand{\be}{\begin{equation}}
\newcommand{\ee}{\end{equation}}
\newcommand{\bea}{\begin{eqnarray}}
\newcommand{\eea}{\end{eqnarray}}
\newcommand{\beaa}{\begin{eqnarray*}}
\newcommand{\eeaa}{\end{eqnarray*}}

\newcommand{\e}{{\rm e}}

\begin{document}

\tolerance=5000

\title{The oscillating dark energy and cosmological Casimir effect}

\author{Olesya Gorbunova$^{1}$\footnote{E-mail:gorbunovaog@tspu.edu.ru}, Diego S\'{a}ez-G\'{o}mez$^{2}$\footnote{E-mail: saez@ieec.uab.es}}


\affiliation{$^{1}$Tomsk State Pedagogical University, Tomsk, Russia}

\affiliation{$^{2}$Institut de Ciencies de l'Espai (IEEC-CSIC), E-08193 Bellaterra
(Barcelona), Spain}

\begin{abstract}

The role of dynamical cosmological Casimir effect to phantom (constant $w$) and oscillating universe is discussed. It is shown explicitly that its role is not essential near to Big Rip singularity. However, the account of Casimir fluid makes the scale factor approach to Rip time to be faster. Rip time itself maybe changed too.

\end{abstract}
\pacs{98.80.-k, 95.36.+x}

\maketitle

\section{Introduction.}

The Dark Energy problem is one of the main task for theoretical physics of last years. Since it was suggested its existence in 1998, a lot of candidates have been proposed (for a review see \cite{DEreview}) in order to give a  natural explanation of what we observe. One of these proposals deals with dark fluids, which possess an inhomogeneous  Equation of State (EoS)  that may depend on time (see Ref. \cite{1}-\cite{3}). In some works, the EoS  is assumed to be periodic in time such that it could explain the so-called coincidence problem (see \cite{NojOdint06}-\cite{OscillCos} and references therein). On the other hand, as some analysis of the observations suggest (Ref. \cite{Observations}), the EoS parameter $w$ for dark energy might have already (or nearly in the future) crossed the phantom barrier ($w<-1$), which could imply some kind of future singularity (for a classification of future singularities see \cite{Sing}). In the current work we study a constant and an oscillating dark energy fluid that crosses this barrier and drives the Universe evolution to a Big Rip singularity. We analyze the implications that a Casimir term could produce on the evolution, as it was pointed in Ref.\cite{brevikDiegoOG} and \cite{Porto}, this Casimir effect could play an important role on the dynamics of the Universe.  Then, the effects on the cosmological evolution and specially close to Big Rip singularity are explored.

\section{The review of the oscillating dark energy universe}

First of all, we review the oscillating universe scenario (see \cite{NojOdint06}, \cite{DSG} and refs.therein).

We consider the universe filled with the ideal fluid (dark energy)
where
the equation of state (EoS) parameter $w$ is time-dependent:
\be
\label{T1}
p=w(t)\rho\ .
\ee

 The conservation law is:
\be
\label{T2}
\dot\rho + 3H\left(p+\rho\right)=0\ ,
\ee
and  the FRW equation
\be
\label{T3}
\frac{3}{\kappa^2}H^2 = \rho\ ,
\ee
where $H=\frac{\dot{a}}{a}$, with $a(t)$ the scale factor. The combination of (\ref{T2}) and (\ref{T3}) gives
\be
\label{T4}
\dot\rho + \kappa\sqrt{3}\left(1+w(t)\right)\rho^{3/2}=0\ ,
\ee
which can be integrated as
\be
\label{T5}
\rho=\frac{4}{3\kappa^2\left(\int dt \left(1+w(t)\right)\right)^2}\ .
\ee
Using (\ref{T4}), the Hubble rate may be found
\be
\label{T6}
H=h(t)\equiv \frac{2}{3\int dt \left(1+w(t)\right)}\ .
\ee
When $w$ is a constant (not equal to phantom divide value), the standard
expression is recovered
\be
\label{T7}
H=\frac{2}{3\left(1+w\right)\left(t-t_s\right)}\ .
\ee
If $\int dt \left(1+w(t)\right)=0$, $H$ diverges,
which corresponds to the Big Rip type singularity,
which occurs when $t=t_s$.
 In case $w>-1$, for the expanding universe
where $H>0$, the cosmological time $t$ should be restricted to be $t>t_s$.

The situation where $w(t)$ is time-dependent, periodic function may be of
physical interest, as it is pointed in several works (see \cite{NojOdint06} and \cite{DSG}) .
An interesting example is given by
\be
\label{T8}
w= -1 + w_0 \cos \omega t\ .
\ee
Here, $w_0,\omega>0$.

Then Eq.(\ref{T8}) gives
\be
\label{T9}
H = \frac{2\omega}{3\left(w_1 + w_0 \sin \omega t\right)}\ .
\ee
Here $w_1$ is a constant of the integration. When $|w_1|<w_0$, the denominator can vanish,
which corresponds to the Big Rip singularity. When $|w_1|>w_0$, there does
not occur such
a singularity.
Since
\be
\label{T10}
\dot H = - \frac{2\omega^2 w_0 \cos \omega t}{3\left(w_1 + w_0 \sin \omega t\right)^2}\ ,
\ee
when $w_0 \cos \omega t<0$ $\left(w_0 \cos \omega t>0\right)$,
the universe lives in phantom
(non-phantom) phase where $\dot H>0$ $\left(\dot H<0\right)$.

\section{The account of Casimir fluid in oscillating and phantom cosmology}

A simple and  natural way of dealing with the
Casimir effect in cosmology is to relate it to the single length
parameter in the ($k=0$) theory, namely the scale factor $a$. It
means effectively that we should put the Casimir energy $E_c$
inversely proportional to $a$. This is in accordance with the
basic property of the Casimir energy, viz. that it is a measure of
the stress in the region interior to the "shell" as compared with
the unstressed region on the outside. The effect is evidently
largest in the beginning of the universe's  evolution, when $a$ is
small. At late times, when $a\rightarrow \infty$, the Casimir
influence should be expected to fade away. As we have chosen $a$
nondimensional, we shall introduce an auxiliary length $L$ in the
formalism. Thus we adopt in model in which
\begin{equation} E_c=\frac{C}{2La}, \label{13}
\end{equation}
where $C$ is a nondimensional constant. This is the same form as
 encountered for the case of a perfectly conducting shell. In the last-mentioned case, $C$ was found to
 have the value
\begin{equation}
C=0.09235. \label{14}
\end{equation}

In the following we shall for definiteness assume $C$ to be
positive, corresponding to a repulsive Casimir force, though $C$
will not necessarily be required to have the value (\ref{14}). The
repulsiveness is a characteristic feature of conducting shell
Casimir theory, following from electromagnetic field theory under
the assumption that dispersive short-range effects are left out. Another assumption
that we shall make, is that $C$ is small compared with unity. This
is physically reasonable, in view of the conventional feebleness
of the Casimir force.
The expression (\ref{13}) corresponds to a Casimir pressure
\begin{equation}
p_c=\frac{-1}{4\pi (La)^2}\frac{\partial E_c}{\partial
(La)}=\frac{C}{4\pi L^4a^4}, \label{15}
\end{equation}
and leads consequently to a Casimir energy density $\rho_c \propto
1/a^4$.

Repeating steps of \cite{brevikDiegoOG} we have:

\begin{equation}
p_c=\frac{C}{8\pi L^4a^4}, \quad \rho_c=\frac{3C}{8\pi L^4a^4}.
\label{18}
\end{equation}

Than, FRW equation of motion looks as

\begin{equation}
\frac{3}{\kappa^2} H^2=\rho+\frac{C_1}{a^4},
\label{eqmotion}
\end{equation}

as Eq.(\ref{T2}). Note that combination of two non-coupled fluids maybe considered as single inhomogeneous fluid (see \cite{1} and \cite{2}).

The conservation law Eq.(\ref{T2}) is the same, because fluids do not couple:

\begin{equation}
\dot{\rho}+3H(p+\rho) =0. \label{EnCons}
\end{equation} 

Substituting (\ref{eqmotion}) into (\ref{EnCons}) one gets:

\begin{equation}
\dot{\rho}+\kappa\sqrt{3}(1+w(t))\rho \sqrt{(\rho+\frac{C_1}{a^4})} =0. \label{think}
\end{equation}

In general, this equation is very difficult to be solved analytically. Although, by deriving the Friedmann equation (\ref{eqmotion}) respect $t$ and combining with the energy conservation equation (\ref{EnCons}), it yields
\begin{equation}
\dot{\theta}+\frac{1}{2}(1+w(t))\theta^2=- (1-3w(t))\frac{3GC}{2L^4a^4}\ ,
\label{D.1}
\end{equation}  
where $\theta(t)=3H(t)$. The case of viscous dark fluid \cite{brevikDiegoOG} is very similar. This equation describes the dynamics of the Universe evolution for a given function $w(t)$, and it is solved as $a=a(t)$. It is easy to see that when the equation of state parameter $w=1/3$, the contribution of the Casimir effect to the dynamics of the Universe vanishes. Let us analyze first of all, a simple example when $w(t)=w_0$ is a constant. For this case, the equation (\ref{D.1}) is rewritten as
\be
\ddot{a}+\frac{1+3w_0}{2}\frac{\dot{a}^2}{a}=-(1-3w_0)\frac{GC}{2L^4a^3}\ .
\label{D.2}
\ee
And the general solution for this equation results:
\be
t-t_s=\int \frac{ada}{\sqrt{ka^{1-3w_0}+\frac{GC}{L^4}}}\ ,
\label{D.3}
\ee
where $k$ is an integration constant. When we have a phantom field with $w_0<<-1$, the solution takes the form:
\be
a(t)=B\left[\left(1+\frac{2}{\e^{\sqrt{kB}(t_s-t)}-1} \right)^2 -1  \right]^{1/1-3w_0}\ ,
\label{D.4}
\ee 
here $B=\frac{GC}{kL^4}$. Note that for $t\rightarrow t_s$, the scale factor given by ($\ref{D.4}$) diverges, such that the so-called Big Rip singularity takes place. On the other hand, if we neglect the Casimir contribution $0<C<<1$, the solution for the scale factor yields
\be
a(t)\sim (t_s-t)^{\frac{2}{3}(1+w_0)}=(t_s-t)^{-\frac{2}{3}|1+w_0|}\ ,
\label{D.5}
\ee
which is the solution in absence of the Casimir term (\ref{T7}). This solution grows slower than in the above case when the Casimir contribution is taken into account (\ref{D.4}). Also note that  the Rip time $t_s$ for each case is different if the Casimir contribution is neglected in the second case, due to $t_s$ depends on the content from the Universe as it can be seen by the first Friedmann equation $t_s-t_0=\int^{\infty}_{a_0} \frac{da}{H_0 a(\sum_i \Omega^i(a))}$, where $t_0$ denotes the current time and $\Omega^i(a)$ is the cosmological parameter for the component $i$ of the Universe  \\
Let us now to analyze the equation (\ref{D.1}) when $w=w(t)$  is a periodic function (\ref{T8}). By inserting  (\ref{T8}) into (\ref{D.1}), the equation turns much more complicated than the constant case studied above. As the Casimir contribution is assumed to be very small, we can deal the equation (\ref{D.1}) by perturbation methods. Hence, the solution might be written as 
\be
\theta(t)= \theta_0 +C\theta_1 + O(C^2)\ .
\label{D.6}
\ee
And by inserting (\ref{D.6}) into (\ref{D.1}), the zero and first order in perturbations can be separated, and it yields the following two equations:
\[
\dot{\theta_0}+\frac{1}{2}w_0\cos\omega t \theta_0=0\ , 
\]
\be
\dot{\theta_1}+w_0\cos\omega t \theta_0 \theta_1=-\frac{3G}{2L^4}(4w_0\cos\omega t)\frac{1}{a^4_0(t)}\ , 
\label{D.7}
\ee
where $a_0(t)=\exp\left(\int dt\frac{2\omega}{3(w_1+w_0\sin\omega t)} \right)$ is the solution for the zero order calculated in the previous section, see (\ref{T9}). We suppose for simplicity $w_1 \sim w_0$, such that a Big Rip singularity appears when $1+\sin\omega t_s=0$. As we are interested in the possible effects close to the singularity, we can expand the trigonometric functions  as series around $t_s$. Then, the second equation in (\ref{D.7}) takes the form: 
\[
\dot{\theta_1}+\frac{4}{3\omega^2}\frac{1}{t_s-t}\theta_1=
\]
\be
=-\frac{3G}{2L^4}(4+\frac{3w_0}{\omega}(t_s-t)) \e^{-\frac{16}{3\omega (t_s-t)} }\ ,
\label{D.8}
\ee
whose solution is given by
\[
\theta_1(t)= k(t_s-t)^{\frac{4}{3\omega^2}}+(t_s-t)^{\frac{4}{3\omega^2}}g(1/t_s-t)\ ,
\]
where
\be
g(x)=\frac{3GC}{2L^4B}\left[ 4x^A\e^{-Bx}- \frac{4(4A+B)}{B}\int x^{A-1}\e^{-Bx} dx \right] \ .
\label{D.9}
\ee
Here $A=\frac{4-6\omega^2}{3\omega^2}$ and $B=\frac{16}{3\omega}$. Then, we can analyze the behavior from the contribution to the solution at first order (\ref{D.9}) when the Universe is close to the singularity. When $t\rightarrow t_s$ ($x\rightarrow \infty$), the function (\ref{D.9}) goes to zero, what means that the Casimir contribution has no effect at the Rip time.

Thus, we have  demonstrated, that dynamical Casimir effect gives no essential contribution to (phantom oscillating) dark energy dynamics near to Rip singularity. 

\section{Discussion}

We discussed the role of dynamical cosmological Casimir effect to phantom (constant $w$) and oscillating universes. It is shown explicitly that its role is not essential near to Big Rip singularity. Moreover, the Rip time $t_s$ may change due to the account of Casimir fluid contribution. For oscillating universe approaching to finite-time singularity, the account of dynamical Casimir effect is made approximately, using perturbations technique. It is possible that only quantum gravity account \cite{5} may prevent the occurrence of Big Rip singularity.

\section*{Acknowledgments}
We are grateful to I.Brevik and S.D. Odintsov for stimulating discussions.
 O. Gorbunova acknowledges support from the GCOE Programme of Nagoya University, the ESF Programme "New
Trends and Applications of the Casimir Effect" and by the Grant for LRSS Project N.2553.2008.2. and thanks Prof. S.Nojiri for kind hospitality. D. S\'aez-G\'omez acknowlegdes a grant from MICINN (Spain), project FIS2006-02842.


\begin{thebibliography}{99}

\bibitem{DEreview}Copeland EJ, Sami M, Tsujikawa S. Dynamics of dark energy. Int J Mod Phys D 2006; 15:1753.
\bibitem{1} Nojiri S, Odintsov SD. Inhomogeneous equation of state of the universe: Phantom era, future singularity, and crossing the phantom barrier. Phys Rev D 2005; 72:023003. 
\bibitem{2}Capozziello S, Cardone V, Elizalde E, Nojiri S, Odintsov SD. Observational constraints on dark energy with generalized equations of state. Phys Rev D 2006; 73:043512.
\bibitem{3} Nojiri S, Odintsov SD. Introduction to Modified Gravity and Gravitational Alternative for Dark Energy. Int J Geom Meth Mod Phys 2007; 4:115-146.
\bibitem{NojOdint06} Nojiri S, Odintsov SD. The oscillating dark energy: future singularity and coincidence problem. Phys Lett B 2006; 637:139-148.
\bibitem{DSG}S\'aez-G\'omez D. Oscillating Universe from scalar-tensor theory and inhomogeneous EoS dark energy. Grav Cosmol 2009; 15:134. 
\bibitem{OscillCos}Hua-Hui Xiong, Yi-Fu Cai, Taotao Qiu, Yun-Song Piao, Xinmin Zhang. Oscillating universe with quintom matter. Phys Lett B 2008; 666:212-217.
\bibitem{Observations}Perivolaropoulos L. Accelerating Universe: Observational Status and Theoretical Implications. astro-ph/0601014.
\bibitem{Sing} Nojiri S,  Odintsov SD, Tsujikawa S. Properties of singularities in the (phantom) dark energy universe. Phys Rev D 2005; 71:063004.
\bibitem{brevikDiegoOG} Brevik I, Gorbunova O, S\'aez-G\'omez D. Casimir Effects Near the Big Rip Singularity in Viscous Cosmology. Preprint gr-qc/arXiv:0908.2882. 
\bibitem{Porto}Herdeiro Carlos AR, Sampaio M. Casimir energy and a cosmological bounce. Class Quant Grav 2006; 23:473.
\bibitem{5}Elizalde E, Nojiri S, Odintsov SD. Late-time cosmology in (phantom) scalar-tensor theory: dark energy and the cosmic speed-up.  Phys Rev D 2004; 70:043539.


\end{thebibliography}
\end{document}